\documentclass[a4paper,11pt]{article}
\usepackage{amssymb,amsmath, amsthm}
\usepackage{epsfig}
\usepackage[english]{babel}
\usepackage{latexsym}
\usepackage{amscd}
\usepackage{braket}
\usepackage{esint}
\usepackage{yfonts}
\usepackage{amssymb}
\usepackage{slashed}
\usepackage{lmodern,textcomp}

\theoremstyle{definition}

\begin{document}

	\title{A  model for stocks  dynamics\\
	based on a non-Gaussian path integral}
	\date{}
	\author{Giovanni Paolinelli \and Gianni Arioli}
	\maketitle
	
	\begin{center}
		{\small Dipartimento di Matematica}\\
		{\small Politecnico di Milano }\\
		{\small piazza Leonardo da Vinci 32, 20133 Milano, Italy}\\
		\texttt{\small giovanni.paolinelli@polimi.it, gianni.arioli@polimi.it}
		\end{center}{\small \par}
	
\begin{abstract}
We introduce a model for the dynamics of stock prices based on a non quadratic path integral.
The model is a generalization of Ilinski's path integral model, more precisely we choose a different action, which can be tuned to different time scales.
The result is a model with a very small number of parameters that provides very good fits of some stock prices and indices fluctuations.
\end{abstract}


	\section{Introduction}
	Stochastic calculus has been a fundamental tool in finance since the publication of the Ph.D. thesis of Louis Bachelier \cite{Bc} in 1900. Much more recently, alternative approaches based on the techniques used in quantum mechanics and quantum field theory have been proposed. In particular, the path integral approach to quantum mechanics introduced by Feynman in \cite{Fey1,Fey} has turned out to be particularly effective for financial applications, see also [4-12]. 
	
	Ilinski in $\cite{Ila,Il}$ establishes two analogies between finance and  quantum mechanics: arbitrage is compared with the Lagrangian action, and the invariance of the behaviour of the market with respect to the choice of the currency of reference is seen as a gauge invariance.
	
	In classical mechanics, the evolution of a dynamical system can be described
    as the trajectory that minimizes the Lagrangian action. In quantum mechanics one can describe the evolution of a system only in probabilistic terms, but the least action principle is not completely lost: loosely speaking, the transition probability between different states can be computed with the aid of a path integral by considering all trajectories, assigning a probability depending on the action to each trajectory and averaging among all trajectories. Such probability is maximal for the least action trajectory, and Plank's constant $h$ determines how fast the probability decreases when other trajectories are considered.

	Classical finance is based on a condition which resembles the principle of least action: arbitrage, i.e., the possibility to make a positive amount of money without risk, is minimized by the market, the minimum being zero.
	
	Ilinski proposes a new approach where arbitrage behaves as the action in quantum mechanics. The most probable trajectory for a stock price in the financial market is the one corresponding to zero arbitrage, but trajectories with some amount of arbitrage are possible, although their probability decreases exponentially with respect to the arbitrage.
	
	The second analogy proposed by Ilinski is gauge invariance. The behaviour of the investors, and hence the trajectory of the prices in a market, does not depend on the choice of currency. Then, all quantities used in the theory have to be invariant under the action of a Lie group $\mathbb{R^+}$, which represent the change between currencies.
	
	Ilinski describes the financial market as a principal fibre-bundle (PFB) with gauge group $\mathbb{R^+}$. This concept is borrowed from physics as well, and it is proposed as the best way to generate the gauge-invariant quantities, since principal fibre-bundles are specifically designed to describe environments where gauge-invariance is required.
	
	The  model described in $\cite{Ila}$ corresponds to a Geometric Brownian Motion (GBM), which generates a normal probability density function; this is not satisfactory, since stocks prices are known to display a leptokurtic distribution. Large price variations are  underestimated by the GBM, causing significant errors in the computation of quantities depending on the stock price trajectories, such as the price of derivatives. More precisely, in \cite{P-D} it is shown that, if the fat tails are ignored, then the rate of return of an investment with no risk of financial loss and the term premium (the compensation that investors require for bearing the risk that short-term Treasury yields do not evolve as they expected) are miscomputed.
 
	Ilinski also introduces in $\cite{Il}$  an improved model combining the minimization of the arbitrage with a perturbation generated by the impact of the orders.
	
	This model has been numerically studied and improved in $\cite{PA}$, where the numerical computations show a good agreement between the theoretical model and the real data concerning the short term dynamics.
	
	The results of $\cite{PA}$ clearly show that the perturbation plays a fundamental role in the good agreement between the model and the real market, since it is responsible of the change of the distribution of prices from mesokurtic to leptokurtic. A similar approach has been developed in \cite{D-F,D-F2,D-F3}, where a model based on quantum field theory has been developed. 
	
	All these results imply that the minimization of the arbitrage does not suffice to describe the short term dynamics of the financial market.
	
	The perturbation of the GBM is used also in classical finance, with stochastic volatility models; this approach has the main drawback of a high number of parameters.
	In this paper we propose to maintain the basic ideas introduced by Ilinski, but we choose a different action. In the following sections, the interpretation of the financial market as principal fibre bundle is assumed, and we refer to $\cite{Ila,Il}$ for an introduction of the topic.
	
	\section{The perturbed models}
	The Heston model is a generalization of the GBM, where the variance is also described by a stochastic process. It consists of the following system of stochastic differential equations:
	
	\[dS_t=\mu S_tdt+\sigma_tS_tdW_t^{(1)},\]
	\begin{equation}\label{stva}v_t=\sigma_t^2,\end{equation}
	\[dv_t=-\gamma(v_t-\theta)dt+k\sqrt{v_t}dW_t^{(2)},\] 
	where $\mu, \gamma,\theta$, $k$  are constants. $W^{(2)}_t$ and $W^{(1)}_t$  are standard Wiener processes correlated by
	\[
	dW_t^{(2)}=\rho dW_t^{(1)}+\sqrt{1-\rho^2}dZ_t \quad \textup{with} \quad \rho \in [-1,1],
	\]
	and $Z_t$ is a Wiener process independent of $dW_t^{(1)}$.
	 
    The stochastic behaviour of $\sigma$ is determined by the parameter $k$; if  $k$ is equal to zero the variance is a deterministic quantity, and the process is a GBM with time dependent variance.
    In \cite{He} also, where this model is used to describe the dynamics of the Dow-Jones index, the importance of the perturbation is stressed.
	
	The approach proposed by Ilinski in \cite{Il}, combining the minimization of the arbitrage with the orders, has similarities with the Heston model. It assumes a closed environment, where the total capital $M$ is constant. The initial allocation of the portfolio is described by the pair $(n_1,m_1)$, where $n_1$ stands for the portion  of wealth invested in cash and $m_1$ is the portion invested in shares, while $(n,m)$  represents the portfolio allocation  at the final time. Clearly $n_1+m_1=n+m=M$. The model simulates all the possible trading patterns that link the initial state with the final. Each trading pattern is composed of a series of buy and sell orders that perturb the main Gaussian dynamics. 

    The improved model proposed in \cite{PA} evaluates the transition probability $P(S(T),(n,m)|S(0),(n_1,m_1))$ from the initial state $(S(0),(n_1,m_1))$ to the final one $(S(T),(n,m))$ with  the following path integral
    \begin{align}\label{comod}
&P(S(T),(n,m)|S(0),(n_1,m_1))=\frac1{n!m!}S(T)^{-\tilde{\beta}\frac{(n-m)}{2}}S(0)^{\tilde{\beta}\frac{(n_1-m_1)}{2}}\\
&\times\int d\bar{\psi}d\psi I(\bar{\psi},\psi,S(0),S(T))\bar{\psi}_{1,0}^{n_1}\bar{\psi}_{2,0}^{m_1}\psi_{1,N}^{n}\psi_{2,N}^{m}e^{-\sum_{j=1,2}\psi_{j,0}\bar{\psi}_{j,0}+\psi_{j,N}\bar{\psi}_{j,N}}\,,\nonumber
\end{align}
where
\[\int d\bar{\psi}d\psi=\prod_{k=1,2}\prod_{i=0,N}\int \frac{1}{2\pi i}d\bar{\psi}_{k,i}d\psi_{k,i}\,,\]
\[I(\bar{\psi},\psi,S(0),S(T))=\int D\log(S)D\bar{\psi}_1D\psi_1D\bar{\psi}_2D\psi_2e^{s_1}\,,\]
and
\begin{equation}\label{act}
s_1=-\frac{1}{2\sigma^2}\int_0^T\bigg[\frac{d}{dt}\bigg( \log(S)-tr-\sum_{j=1}^{N}\alpha_j(||\psi_2||^2-||\psi_1||^2)\bigg| ||\psi_2||^2-||\psi_1||^2 \bigg|^{\Gamma_j-1} \bigg)\bigg]^2dt\end{equation}\[+
\int_0^T\bigg(\frac{d\bar{\psi}_1}{dt}\psi_1+\frac{d\bar{\psi}_2}{dt}\psi_2+\Delta^{-1}S^{\tilde{\beta}}\bar{\psi}_1\psi_2+\Delta^{-1}S^{-\tilde{\beta}}\bar{\psi}_2\psi_1\bigg)dt\,.\nonumber
\]
The model proposed by Ilinski in \cite{Il} is a special case of the model in \cite{PA}, with $N=1$ and $\Gamma_1=1$. More precisely, the formulas above represent the continuous time version of the model described in $\cite{PA}$.\\
We recall that the result of the previous integral is a complex number and the probability is its square module. We maintain the same notation of \cite{Il}, \cite{PA} in order to simplify the comparison.

The first integral in the definition of the action $s_1$ in (\ref{act}) determines the stock price dynamics, while the second integral determines the portfolio dynamics.
The connection between the two dynamics is controlled by the parameters $\alpha_j$.  As in the Heston model, if $\alpha_j=0$ for all $j$, then the previous integral becomes 
\begin{equation}\label{pgbm} P(S(T)|S(0))_{GBM} \propto \int D\log(S)e^{-\frac{1}{2\sigma^2}\int_{0}^{T}(\partial_t\log(S_t)-r)^2dt},
\end{equation} 
which is the GBM formulation proposed by Ilinski in \cite{Ila}. We observe that the previous model shares the same ideas of (\ref{stva}), since it represents a perturbation of the GBM. Heuristically, the parameters $\alpha_j$ correspond to the parameter $k$ in the Heston model.

Similar results has been achieved by Dupoyet et. al. in \cite{D-F,D-F2,D-F3}, where the GBM model is perturbed using an extra term derived by a quantum lattice model.

The common background of all the previous models is a Gaussian dynamics perturbed with an extra term. This approach requires the choice of many parameters and the computation of many variables to control the strength of the perturbation. The parameters can be split in three classes:
\begin{itemize}
	\item Parameters of the main (Gaussian) dynamics.
	\item Parameters determining the dynamics of the perturbation.
	\item Parameters determining the coupling between the main dynamics and the perturbation. 
\end{itemize} 
As we pointed out, the perturbation is essential, since the Gaussian dynamics alone does not fit the market behaviour. The main idea that we present in this paper is a model where the main dynamics is not Gaussian, so that a perturbation is not required, and only parameters for the main dynamics are introduced.
Our goal  is a model with a precision similar to the model in \cite{PA,D-F,He}, but simpler and with a lower number of parameters.

\section{The non-Gaussian path integral}
We begin by observing how the path integral encodes the Gaussian behaviour. Consider the stochastic differential equation
\begin{equation}
\dot{x}(t)=\eta(t)\,
\end{equation}
where
\[
x(t)=\log(S_t).
\]
The noise $\eta$ does not need to be Gaussian; let $\tilde{D}(\eta)$ be the probability distribution of $\eta$ and define $L(\eta)$ by
\begin{equation}\label{lr}
\tilde{D}(\eta)=e^{-L(\eta)}.
\end{equation}
To underline the analogy with quantum mechanics, we call the function $L(\eta)$ the Lagrangian.
In \cite[see 20.1.13]{HKl} it is shown that the path integral
\begin{align}
P(x_b,t_b|x_a,t_a)&=\int D\tilde\eta\int Dx \exp[-\int_{t_a}^{t_b}L(\tilde\eta(t))dt]\delta(\dot{x}-\tilde\eta)\notag\\
&=\int Dx \exp[-\int_{t_a}^{t_b}L(\dot{x})dt]\label{gpi}
\end{align}
returns the probability distribution of $x_b=x(t_b)$ when $x(t_a)=x_a$.
Comparing (\ref{pgbm}) with (\ref{gpi}), we see that the Lagrangian associated to the GBM described by Ilinski is
\begin{equation}\label{lag}L_{GBM}=\frac{1}{2\sigma^2}(\partial_t\log(S_t)-r)^2 \end{equation}
and then
\begin{equation}\tilde{D}(\partial_t\log(S_t))\simeq N(r_{GBM},\sigma^2)\,\end{equation}
with
\[r_{GBM}=r-\sigma^2 T/2\,.\]
The previous analysis, together with  formula (\ref{lag}), explains why  the log-returns in (\ref{pgbm}) are normally distributed and why the Gaussian nature of the perturbation is encoded in the structure of the Lagrangian. The latter observation is the starting point of our model.

{\bf Remark.} Formula (\ref{pgbm}) represents a transition probability of the price $S(t)$ expressed as an integral in $D\log S_t$, while (\ref{gpi}) represents the transition probability of $x=\log(S)$ expressed as an integral in $Dx$.  
This difference is due to the fact that (\ref{gpi}) is not written in an explicit gauge-invariant form, since the measure $D\log(S)=DS/S$ is gauge-invariant ($D\log(S)=DS/S$), while the measure $DS$ is not.

\section{The new model}
\subsection{Setting}
In order to present the basic hypotheses underlying the new model, we begin with the definition of the path-space in the financial context.
Given a time interval $[0,T]$, where $T\in \mathbb{R^+}$ represents our choice of time horizon, we define
\begin{equation}\label{psc}
X_{S_0,S_T}=\{S\in BV([0,T],\mathbb{R}^+)\,:\,S(0)=S_0,\,S(T)=S_T \}
\end{equation}
as the set of all the paths connecting two prices $S_0,S_T\in \mathbb{R^+}$.
The set of all admissible paths is
\begin{equation*}
X_T=\bigcup_{S_0,S_T\in\mathbb{R^+}}X_{S_0,S_T}\,.
\end{equation*} 
We consider the BV functions in order to take in account both continuous and non-continuous stock-price paths.
 
The basic hypotheses for our model are:
\begin{itemize}
    	\item\textbf{(HP1)} There exists a functional $\mathcal{A}:X_T\to\mathbb{R}$, invariant with respect to the action of the gauge group $\mathbb{R^+}$, and related to the amount of money that is possible to earn with some trading strategy during the time horizon $[0,T]$.
    	\item \textbf{(HP2)} The market evolves fluctuating around the configuration of minimal value of $\mathcal{A}$.
    	\item \textbf{(HP3)} The financial market is not fully efficient, small amounts of arbitrage are allowed.
   \end{itemize}

\subsection{Stochastic minimization}
In order to state rigorously what we mean by ``fluctuation around the configuration of minimal action",  we introduce the concept of {\em stochastic minimization}.
Assume that $\mathcal{A}:X_T\to\mathbb{R}^+$ is an action which satisfies {\bf HP1}, let
$P:\mathbb{R}^+\to \mathbb{R^+}$ be a decreasing function, and let
\begin{equation}
p:X_T\to\mathbb{R}^+\,,\quad p(S)=P(\mathcal{A}(S)).
\end{equation}

Since $P$ is decreasing, the classical path $\bar S$ which minimizes the action, corresponds to  the path  that maximizes $p$. By {\em stochastic minimization} we are assuming that all paths are possible, with probability assigned by $p$. All quantities related to the path of prices (probability of the final price or related quantities, such as the value of a derivative product) has to be computed on each path, and the result is the weighted average with respect to $p$ on all paths. There are two main differences with respect to the computation of a propagator in quantum mechanics: first, in quantum mechanics the action is multiplied by the imaginary unit and the transition probability is the modulus square of the path integral; second, for the financial path integral we only require that $P$ is a decreasing function, not necessarily an exponential.
The analogy with quantum mechanics also concerns the methods adopted to compute the transition probability, as we show in Section \ref{sec:nm}.
It is straightforward to observe that the system described by Ilinski consists in the {\em stochastic  minimization} of the arbitrage, with respect to an exponential function $P$.

\subsection{The new action}
We denote by $\mathcal{A}_{arb}$  the amount of money that is possible to earn with  arbitrage in the time horizon $[0,T]$.
Ilinski shows in $\cite{Ila}$  that
\begin{equation}\label{arb}
   \mathcal{A}_{arb}(S_t,T)=\beta_2\int_{0}^{T}(\partial_t\log(S_t)-r_t)^2dt=\beta_2\int_{0}^{T}\Omega(t)^2dt,
\end{equation}
   where $r$ is the net non-risky interest rate and $\beta_2$ is a constant with the dimension of time.
   The function $\Omega(t)$ is the curvature of the principal fibre-bundle considered by Ilinski, therefore $\mathcal{A}_{arb}$ is gauge invariant, see \cite[Chap. 5]{Il}. Note that the curvature has the dimension of time inverse, therefore $\beta_2$ needs to have the dimension of time; similarly, in quantum mechanics case the action is divided by $h$. Note that $\beta_2$ weighs the paths variation with respect to $\bar S$, that is it corresponds to the inverse of the variance squared.
   
   Since the curvature is gauge invariant, it is a natural component of a gauge-invariant functional. A different gauge invariant functional is the $L^1_{[0,T]}$-norm of $\Omega(t)$
   \begin{equation}\label{me}
   \mathcal{A}_{me}(S_t,T)=\beta_1\int_{0}^{T}|\partial_t\log(S_t)-r_t|dt=\beta_1\int_{0}^{T}|\Omega(t)|dt.
   \end{equation}
   In this case $\beta_1$  provides a scale for the variation of the paths with respect to $\bar S$.
   Adopting the same point of view of Ilinski, we consider $\mathcal{A}_{me}(S_t,T)$ as the maximal amount of money that is possible to earn in the time horizon $[0,T]$. The strategy to obtain that gain consists in borrowing cash and buying the stock when the stock outperform the cash, and vice versa. This strategy involves some risk, since it is not possible to know in advance what is the right case. It is straightforward to check that the functionals $(\ref{arb})$  and $(\ref{me})$ satisfy the hypotheses \textbf{HP1}.
   Note that
      \[
   \min \mathcal{A}_{me}(S_t,T)=\min \mathcal{A}_{arb}(S_t,T)=0,\quad\iff\quad \partial_t\log(S_t)=r_t
   \]
   and, if $T_2 > T_1$ and $S_1(t)=S_2(t)$ for all $t \in[0,T_1]$ , then
   \[
   \mathcal{A}_{me}(S_1(t),T_1) <\mathcal{A}_{me}(S_2(t),T_2)\quad \textup{and} \quad  \mathcal{A}_{arb}(S_1(t),T_1) <\mathcal{A}_{arb}(S_2(t),T_2).
   \]
   It is also important to note that, if $\beta_1=\beta_2$  and assuming that $S_t$ does not minimize the actions, then
   \begin{equation}\label{ineq}
    \mathcal{A}_{me}(S_t,T) > \mathcal{A}_{arb}(S_t,T)\,.
   \end{equation}
   The previous inequality implies that, if we use the same scale  to measure the revenues from $\mathcal{A}_{me}$ and $\mathcal{A}_{arb}$, then the first strategy is more rewarding than the second one.
   
   This last fact may seem counter-intuitive. We recall that (\ref{ineq}) is computed over the continuous-time version of the process $S_t$. In the real case the stock price does not vary continuously, since a price variation happens only when there is a new order with a different price. The time scale associated to the price variation between two successive orders is called  the "tick-by-tick" variation of the price, and it is the time frame that must be  selected to compute the previous quantities.
   Log-price variations in this context are always very small, implying (\ref{ineq}). The analysis of the probability density functions of the one minute dynamics confirms this statement.
  
   We  now show that $\mathcal{A}_{me}$ is a reasonable choice for the action functional in the case of a short time horizon, whereas  $\mathcal{A}_{arb}$ is more suitable in the case of a long time horizon.
   
\subsubsection{The short time horizon}
   Assuming a closed market, we note that the amount of money that a trader gains corresponds to the loss of all the other traders. The fairest situation is the one without losses, which implies no gain as well.
   
   This happens when the value of the functional $\mathcal{A}_{me}$ is equal to 0. Market configurations with $\mathcal{A}_{me}>0$ are associated with losses among other players and therefore they should be considered less probable.
   
   In the short term dynamics, the minimization of the arbitrage  is in contrast with $\textbf{HP3}$. The functional (\ref{arb}) is proportional to $T$ and it is smaller than $ \mathcal{A}_{me}$, i.e., it is negligible.
  
\subsubsection{The long time horizon} Since the amount of arbitrage is directly proportional to the time horizon $T$, we can assume that in this case it is non negligible. Clearly,  inequality (\ref{ineq}) still holds, but in this case we have to consider that arbitrage does not involve risk. It is reasonable to assume that  investors prefer the arbitrage strategy albeit it is less profitable, since the gain is sure. In this situation, it is essential to minimize the arbitrage in order to prevent the possibility to extract considerable amount of money without risk from the market in order to respect $\textbf{HP3}$.
   
\subsubsection{Intermediate-time horizon}
   To consider an intermediate-time horizon, we assume that it is possible to  modify continuously the action functional from $\mathcal{A}_{arb}$ to $\mathcal{A}_{me}$. Given the structure of the previous functionals, the most natural choice seems to be
   \begin{equation}\label{ap}
   \mathcal{A}_{p}(S_t,T)=\beta_p\int_{0}^{T}|\partial_t\log(S_t)-r_t|^pdt\,, \quad p \in [1,2]\,.
   \end{equation}
   This action may be interpreted as a hybrid strategy which involves both arbitrage and risk trading.
   
   The previous formula is unconventional in the theory of path integrals, where the action is usually quadratic plus some perturbation. The previous definitions have to be intended in a formal sense; their interpretation and numerical evaluation is discussed in section \ref{sec:dta}.
   
   \subsection{The new stochastic dynamics}
   We now apply the concept of  stochastic  minimization to the  action functionals previously defined, giving also the explicit formulation of the probabilities adopted.
   
   The arbitrage strategy proposed by Ilinski can, in principle, be followed by every trader in the market.
   This means that it would be very simple to make a profit with an arbitrage transaction, i.e., to extract  consistent amount of money from the market without risk.
   In order to forbid this eventuality, it is necessary that big arbitrage opportunities are very unlikely.
   This justifies the choice adopted by Ilinski which uses the exponential function.
   
   The short time horizon case is different. The most probable configuration of the market does not admit capital losses, but different traders have different trading experience. We can assume that the more experienced trader will make a profit over the unskilled one; this means that high values of the functional $\mathcal{A}_p$ with $p\in[1,2)$; should be more likely, given the same amounts of arbitrage.
   To account for this, we propose as probability density function
   \[P(\mathcal{A}_p)\propto e^{-\beta_p\mathcal{A}_p^\gamma} \quad \textup{with} \quad \gamma \in (0,1); \]
   to maintain the analogy with the GBM case we define $\beta$ by
   \[
   \beta_p=\frac1{2\sigma^p}.
   \] 
   Note that the parameter $\sigma$ is not the volatility, but it is related to it.  
   Considering  the action $\mathcal{A}_{p}$ with the probability  defined above and {\bf HP2}, the fluctuations of the market around the minimum action configuration turn out to be described by
   \begin{equation}
   P_{\gamma,p}(\mathcal{A}_p)\propto e^{-\frac1{2\sigma^p}\big(\int_{0}^{T}|\partial_t\log(S_t)-r_t|^pdt\big)^\gamma}\,,
   \end{equation} 
   and the GBM is recovered when $\gamma=1$ and $p=2$.
   
   It is interesting to compare the pdf obtained with the GBM against different $(\gamma,p)$ cases. The condition $\gamma < 1$, i.e. the intermediate time horizon,
   implies that small amounts of action $\mathcal{A}_p$ are less likely to happen with respect to  the same amount of arbitrage in the GBM case. This is in agreement with our hypothesis, since by \textbf{HP3} small amount of arbitrage are allowed. Moreover, the arbitrage is not the quantity minimized in the short therm dynamics. On the other side, large values of the action  $\mathcal{A}_p$ are more likely to happen with respect to  the same amount of arbitrage in the GBM case; this is in agreement with the condition that requires the minimization of the arbitrage.
   
   Recalling the path integral formalism introduced by Feynman in $\cite{Fey1}$, we obtain that the transition probability is the gauge-invariant weighted mean with respect to  $P_{\gamma,p}$ over the space of all paths connecting $S(0)$ with $S(T)$, i.e., 
   \[
   P(S(T)|S(0))=\tilde{\sum_{S_t\in X_{S_0,S_T}}}P_{\gamma,p}(\mathcal{A}_p(S_t)),
   \] 
   where the tilde notation indicates that the sum is computed adopting a gauge-invariant measure. The above formula means that the transition probability  $P(S(T)|S(0))$ is composed of the contributions of all the possible paths connecting the initial and the final states. Each path contributes inversely with respect the amount of action associated to it; the magnitude of the contributions is controlled by the probability  $P_{\gamma,p}$. An interesting mathematical problem is to define a way to sum all the paths, respecting the gauge invariance; this will be discussed below.\\ 
   The sum over the path space is   an infinite dimensional integral
   \[
     \int D S=\lim_{N \rightarrow \infty}\prod_{i=1}^{N-1}\int_0^{\infty}dS_i.
   \]
   Note that the  definition above is not gauge invariant.
   Given a  gauge group $G$, a gauge-invariant function $f(x)$ and a measure $d\mu(x)$, we require that
   \[
    \int f(x)d\mu(x)=\int f(gx)d\mu(gx) \quad \forall g \in G.
   \]  
   Such measure is known as the Haar measure of the group. For the group $\mathbb{R^+}$ the Haar measure exists and it is unique, up to a multiplicative constant; it is given by
   \[dS/S=d\log(S)\,.\]
   Finally, the formula for the transition probability is
   \begin{equation}
   \label{fm} P(S(T)|S(0))\propto\int D\log(S)e^{-\frac{1}{2\sigma^p}\big(\int_{0}^{T}|\partial_t\log(S_t)-r|^pdt\big)^\gamma}.
   \end{equation} 

\section{Results}
We  present  some comparisons between real data and our model. The  real data consist in 3 months price-sheets of the  {\em AMAZON}, {\em GENERAL ELECTRIC} 
and {\em APPLE} stocks. To obtain the approximate probability density function associated to the   stock, we follow the method introduced in the reference \cite{Bc}, i.e. we consider a set of  historical data as instances of a stochastic variable. We first compute $X_i$ by
\[ X_i=\log(P(t_i)/P(t_{i-1}))\,, \qquad t_i-t_{i-1}=\tau,  \]  
then we build a histogram of the values $X_i$ with $N$ bins. The histograms shown in the figures below  display the number of counts $\Delta C$ in each histogram bin,   divided by the bin width $\Delta S/N$. The result is then normalized.

The error bars for the real data  are estimated as  $ \sigma_{bins}\Delta S/N$; where  $\Delta S$ is the width of the histogram $x$-bars. In order to compare our computations with the results obtained in \cite{Il,M-S,D-F}, we plot the probability density functions in logarithmic scale.
The error associated to the numerical computation has been estimated and it is omitted since it is negligible with respect to the error over the data. The numerical computations are performed approximating the path integrals with a 10 dimensional integral; details about finite dimensional approximations and the algorithms used to evaluate the integrals can be found in the sections 6 and 7. 

\subsection{Short-term dynamics}
We present here the results for three stocks, {\em AMAZON}, {\em GENERAL ELECTRIC} 
and {\em APPLE}.\par  The data used are the price-sheet from September 20, 2017 until December 20, 2017, with sample frequency 1, 5 and 30 minutes. In this section $r=0$ unless otherwise stated. 

\subsubsection{1 minute} 
\begin{figure}[ht]
	\begin{minipage}[b]{5.75cm}
		\centering
		\centering
		\includegraphics[width=5.75cm]{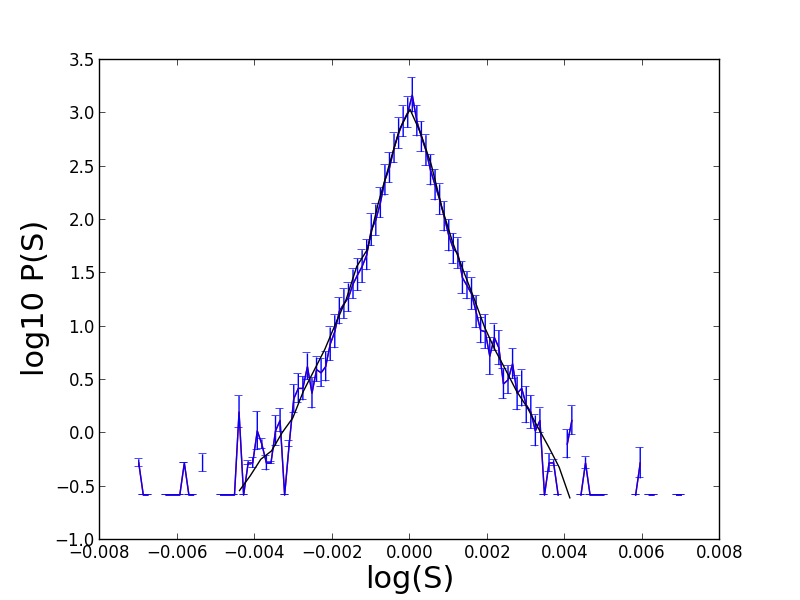}
		\caption{AMAZON stock, $\sigma=0.035$, $p=1.15$, $\gamma=0.15$, $\Delta \log(S)=0.011$, $N=120$ and $T=1m$.}
		\label{am1}

	\end{minipage}
	\ \hspace{2mm} \hspace{3mm} \
	\begin{minipage}[b]{5.75cm}
		\centering
		\includegraphics[width=5.75cm]{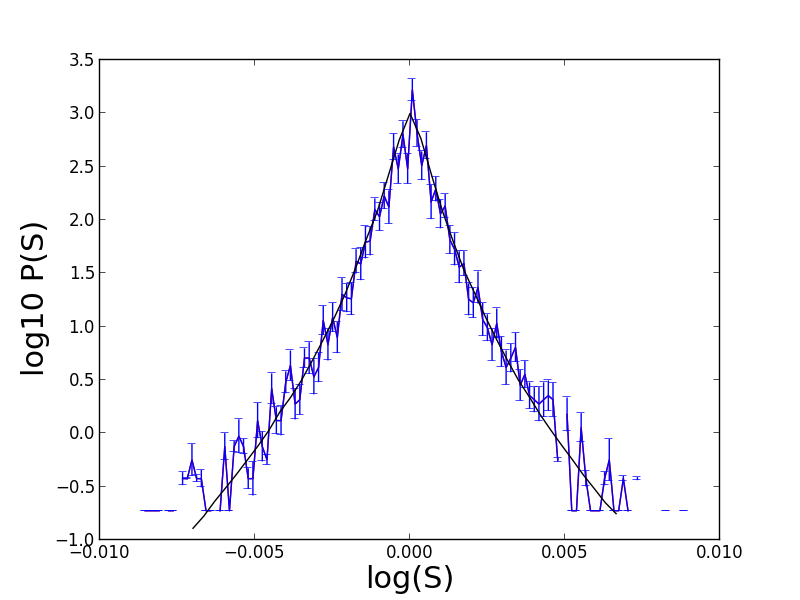}
		\small 
		\caption{GENERAL ELECTRIC Stocks $\sigma=0.0326$, $p=1.15$, $\gamma=0.15$, $\Delta \log(S)=0.014$, $N=120$ and $T=1m$.}
		\label{g1}
	\end{minipage}
\end{figure}

\begin{figure}[ht]
	
	\centering
	
	\includegraphics[width=5.75cm]{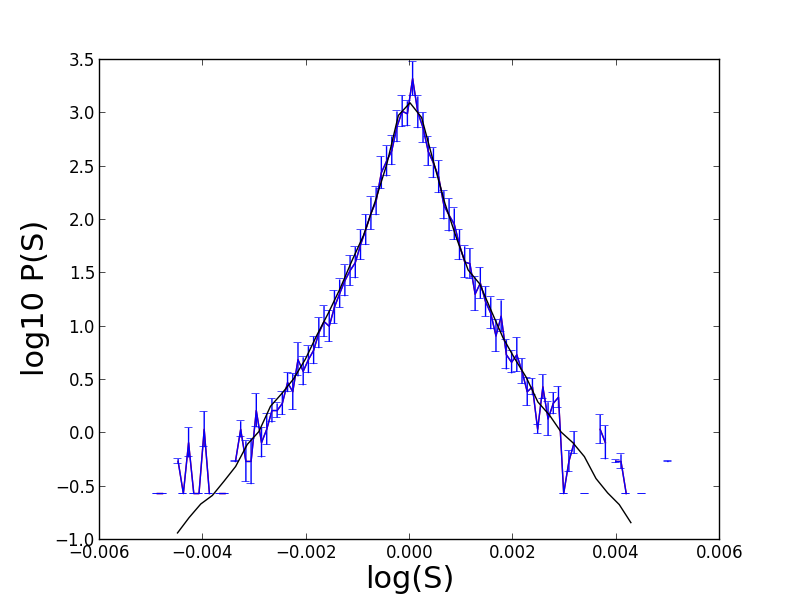}
	\small 
	\caption{  APPLE Stocks $\sigma=0.0321$, $p=1.15$, $\gamma=0.15$, $\Delta \log(S)=0.009$, $N=120$ and $T=1m$.}
	\label{ap1}	
\end{figure}

In this section we present  the results for the one minute dynamics. The  Figures $\ref{am1},\ref{g1}$ and $ \ref{ap1}$ show in blue the probability density function derived from the data, while the black line is the probability density function computed with the model presented above. The dataset consists of 30000 prices.


\subsubsection{5 minutes} 
\begin{figure}[ht]
	\begin{minipage}[b]{5.75cm}
	\centering
		\includegraphics[width=5.75cm]{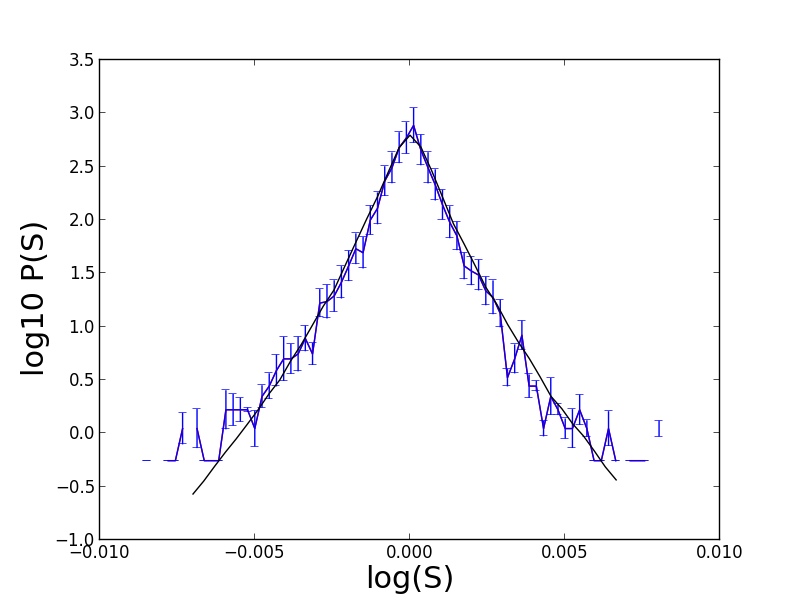}
		\small 
		\caption{ AMAZON stock, $\sigma=0.0119$, $p=1.2$, $\gamma=0.2$, $\Delta \log(S)=0.016$, $N=75$ and $T=5m$.}
		\label{am5}
		
	\end{minipage}
	\ \hspace{2mm} \hspace{3mm} \
	\begin{minipage}[b]{5.75cm}
		\centering
		\includegraphics[width=5.75cm]{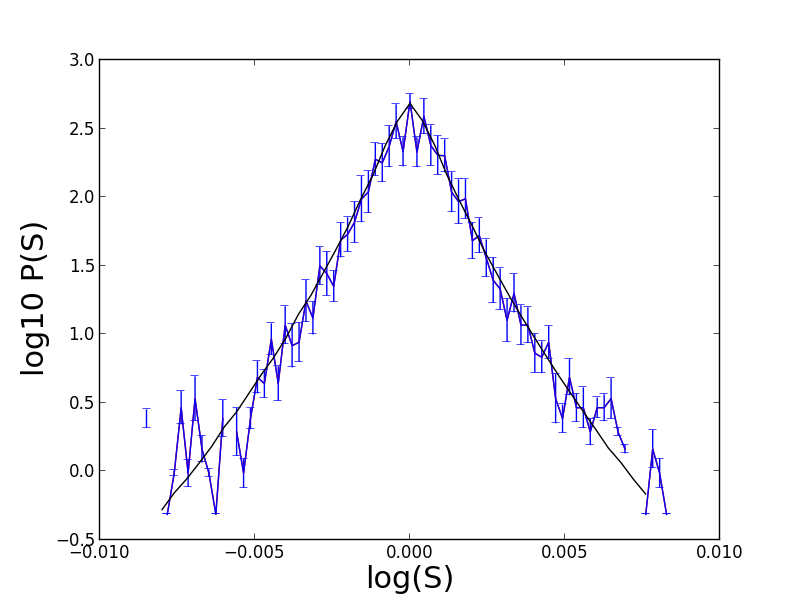}
		\small 
		\caption{GENERAL ELECTRIC Stocks $\sigma=0.0111$, $p=1.2$, $\gamma=0.2$,  $\Delta \log(S)=0.014$, $N=77$ and $T=5m$.}
		\label{g5}
	\end{minipage}
\end{figure}

\begin{figure}[ht]
	
	\centering
	
	\includegraphics[width=5.75cm]{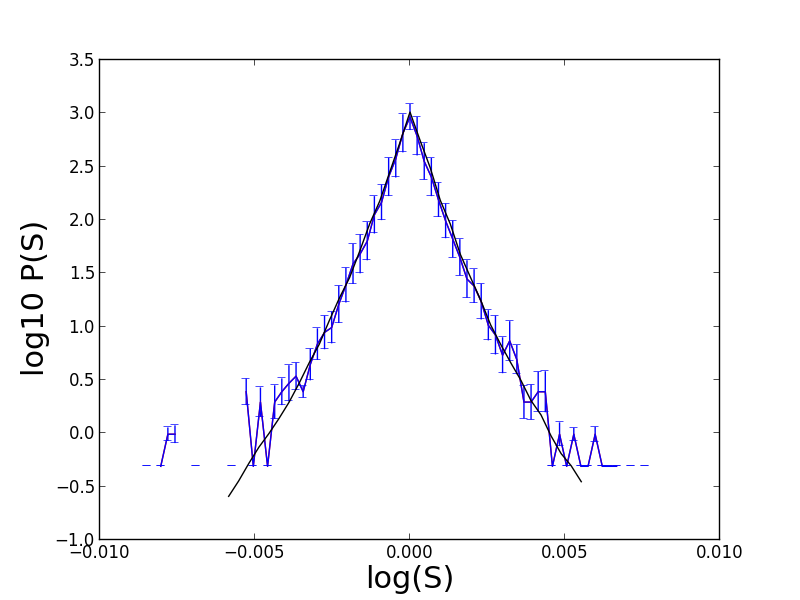}
	\small 
	\caption{ APPLE Stocks $\sigma=0.0107$, $p=1.2$, $\gamma=0.2$, $\Delta \log(S)=0.011$, $N=75$ and $T=5m$.}
	\label{ap5}
\end{figure}

The five minutes dynamics is presented in the Figures $\ref{am5}, \ref{g5}$ and $\ref{ap5}$. The dataset consists of 6000 prices.\par

\medskip
The values of the parameters $\gamma$ and $p$ is the same for  the three stocks.\\
We also note that the shared parameters have a higher values with respect to the one minute dynamics.

\subsubsection{30 minutes } 
\begin{figure}[ht]
	\begin{minipage}[b]{5.75cm}
		\centering
		\includegraphics[width=5.75cm]{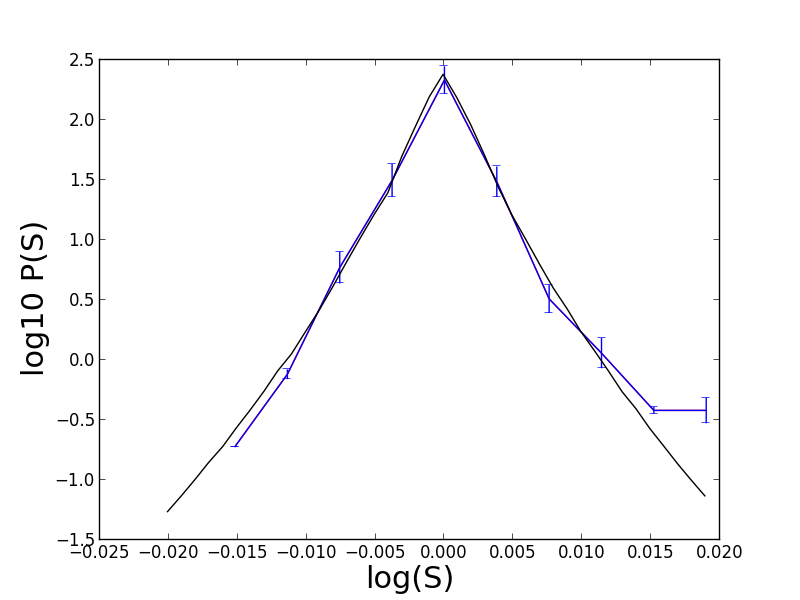}
		\small 
		\caption{ AMAZON stock, $\sigma=0.00945$, $p=1.23$, $\gamma=0.23$, $\Delta \log(S)=0.034$, $N=11$, $r=-0.0001$ and $T=30m$.}
		\label{am30}
		
	\end{minipage}
	\ \hspace{2mm} \hspace{3mm} \
	\begin{minipage}[b]{5.75cm}
		\centering
		\includegraphics[width=5.75cm]{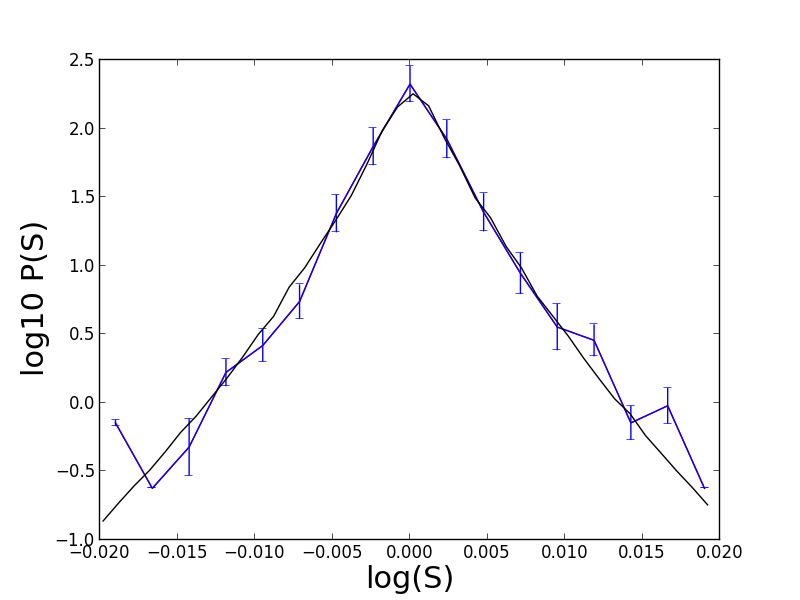}
		\small 
		\caption{GENERAL ELECTRIC Stocks $\sigma=0.0099$, $p=1.23$, $\gamma=0.23$,  $\Delta \log(S)=0.04$, $N=17$, $r=0.0002$ and $T=30m$.}
		\label{g30}
	\end{minipage}
\end{figure}

\begin{figure}[ht]
	
	\centering
	
	\includegraphics[width=5.75cm]{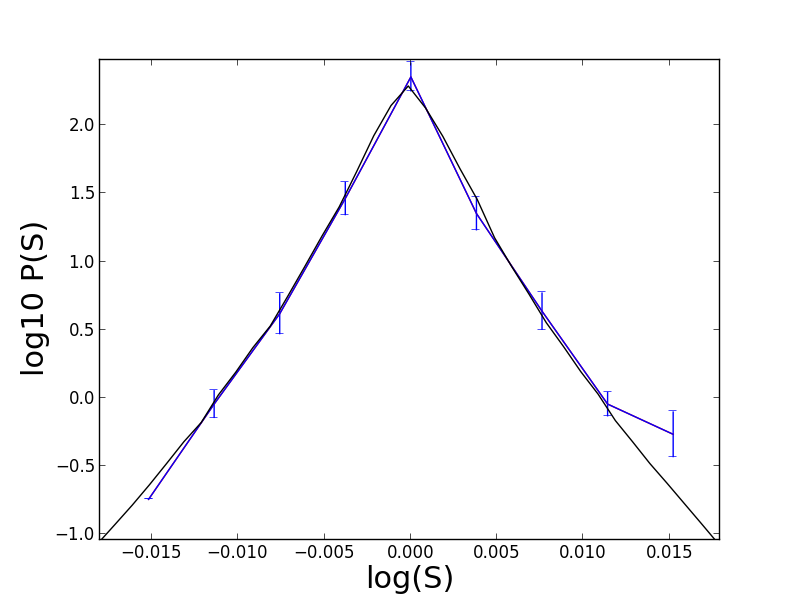}
	\small 
	\caption{ APPLE Stocks $\sigma=0.0094$, $p=1.23$, $\gamma=0.23$, $\Delta \log(S)=0.034$, $N=11$, $r=-0.00015$ and $T=30m$.}
	\label{ap30}
\end{figure}

Finally, we present the final results for the half-hour time frame.
Figures $\ref{am30}, \ref{g30}$ and $\ref{ap30}$ show the results of our computations, together with the probability density functions computed from the real data.
In this case the dataset contains 1000 prices; in order to maintain the number of prices per bin close to the other cases:  $N$  is between 11 and 17.

\subsection{The analysis of the short-term dynamics}
In the previous section we have analysed the dynamics of 3 similar stocks for market capitalization, in 3 different time horizons. The results of the computations shown above are consistent with the  theory. The values of $\gamma$ and $p$ converge to the values associated to the GBM, when the time horizon is increased. It is also interesting to note that the values of the previous parameters are shared among all the stocks analysed for each time horizon treated. This last fact seems to suggest that the three stocks share a common dynamics.

\subsection{Long-term dynamics}
We present here the results with time intervals of one day and one week. In order to have enough data to approximate properly the probability density function, we need a consistent extension of the time horizon, which now is 30 years. We select data from the 
February 01, 1988 until February 01, 2018.\par
Considering such time intervals, it is better to study the dynamics of indices, since stock prices may be affected by capital increases, splits and other events which would require more care in the selection of the data.
We select the {\em Dow Jones} and the {\em S$\&$P500}.

\subsubsection{1 day dynamics}
\begin{figure}[ht]
	\begin{minipage}[b]{5.75cm}
	\centering
\includegraphics[width=5.75cm]{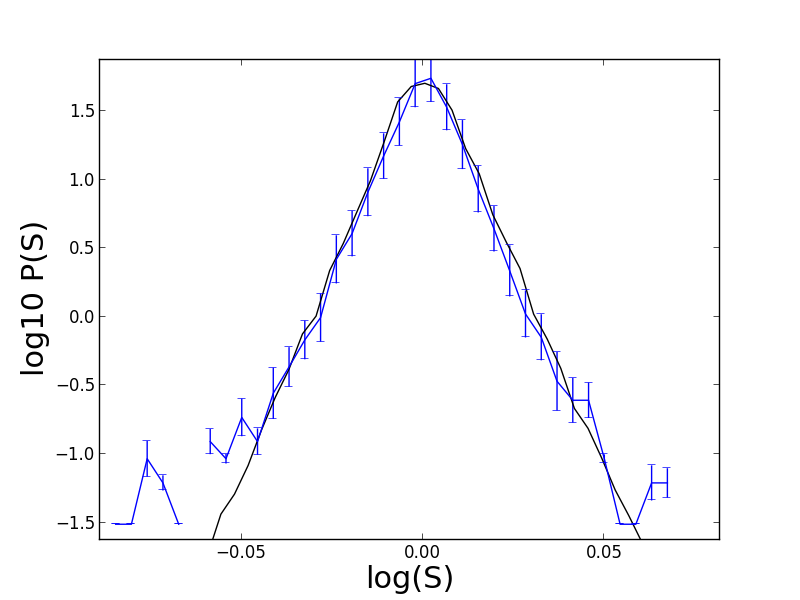}
\small 
\caption{Dow Jones Index, $\sigma=0.062$, $p=1.35$ and $\gamma=0.35$, $\Delta \log(S)=0.12$, $N=70$, $r=0.0005$ and $T=1d$.}
\label{dj1d}

	\end{minipage}
	\ \hspace{2mm} \hspace{3mm} \
	\begin{minipage}[b]{5.75cm}
	\centering
\includegraphics[width=5.75cm]{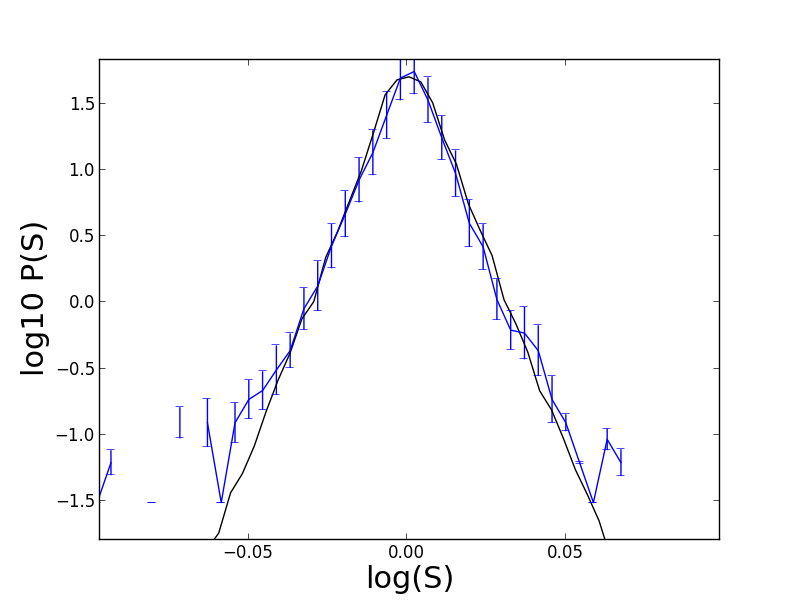}
\small 
\caption{  S\&P500 Index, $\sigma=0.062$, $p=1.35$ and $\gamma=0.35$, $\Delta \log(S)=0.12$, $N=70$, $r=0.0005$ and $T=1d$.}
\label{gspc1d}
	\end{minipage}
\end{figure}

The one day dynamics  is presented in  Figures $\ref{dj1d}$ and $\ref{gspc1d}$.  
The dimension of our dataset is approximatively of 7000 prices, where the price selected is price at the closing time.\par 
	
\subsubsection{ The week  dynamics} 
\begin{figure}[ht]
	\begin{minipage}[b]{5.75cm}
		\centering
		\includegraphics[width=5.75cm]{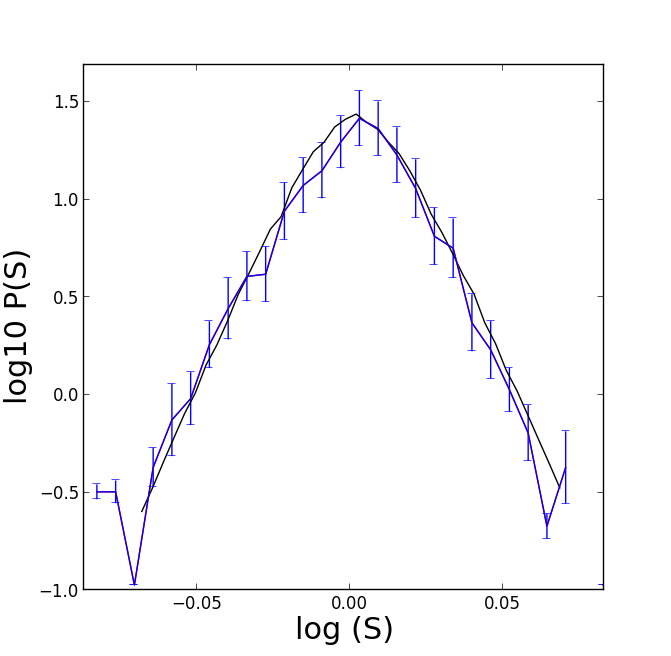}
		\small 
		\caption{ Down Jones Index, $\sigma=0.023$, $p=1.42$ and $\gamma=0.42$, $\Delta \log(S)=0.14$, $T=7d$, $r=0.002$ and  $N=50$.}
		\label{dj1w}

	\end{minipage}
	\ \hspace{2mm} \hspace{3mm} \
	\begin{minipage}[b]{5.75cm}
		\centering
		\includegraphics[width=5.75cm]{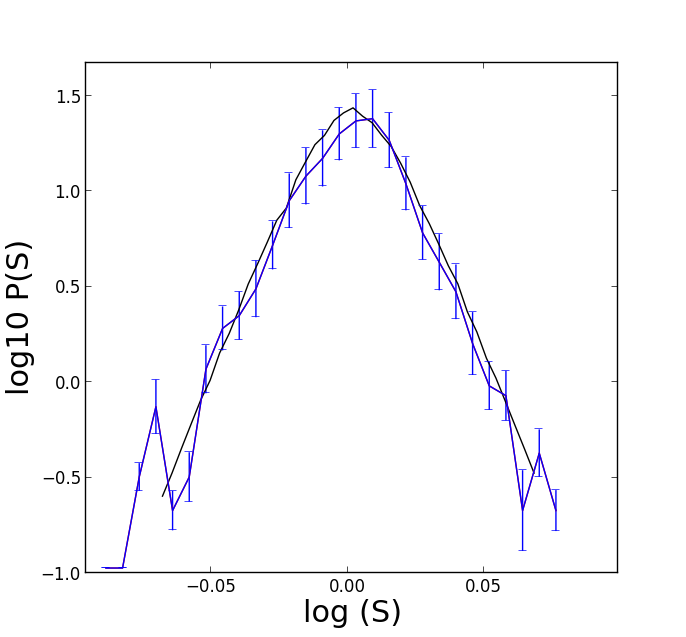}
		\small 
		\caption{  S\&P Index   $\sigma=0.023$, $p=1.42$ and $\gamma=0.42$, $\Delta \log(S)=0.14$, $T=7d$, $r=0.002$ and  $N=50$.}
		\label{gspc1w}
	\end{minipage}

\end{figure}

The results for the weekly dynamics is presented in the Figures $\ref{dj1w}$ and $\ref{gspc1w}$, the data-set contains almost 1100 prices, selected at the Friday closing time. The histograms of these computations are characterized by a lower number of prices per bin. Since we are considering prices with one week sampling frequency, the variance of each bin is very high, therefore we need a larger $N$ to maintain low data errors

\subsection{Final analysis}
The results  with a time horizon of one minute show a better agreement with the data than \cite{Il,M-S,D-F}. Our results are similar to those in \cite{PA}; but we have 3 parameters instead of 5.\\ 
It is also interesting to note that the same model is capable to provide a good agreement with 5 different time horizons in both cases of indexes and stocks.\\  
The parameter values used in the previous computations are consistent with the financial interpretation. We note that the values of $(\gamma,p)$ both grow with the enlargement of the time horizon; this relation is justified by the theoretical background of the theory presented in section 4. 
In all the cases, different stocks and indexes share the same value of the parameters $\gamma$ and $p$. The difference associated to the liquidity is expressed by a different value of $\sigma$. This fact seems to suggest that, given a time frame, there is a common dynamics for all the cases.

\section{Definition and approximation of the path integrals}\label{sec:dta}

There exists a rigorous theory of path integrals only when the Lagrangian is quadratic. When a small perturbation is added, as it happens e.g. in quantum electrodynamics, it is possible to evaluate them thanks to perturbation techniques based on Feynman diagrams, but such techniques are not mathematically rigorous; moreover, they involve infinite quantities which require renormalization theory to be dealt with. In our model, as soon as $p\ne2$, there is no quadratic part to use as a starting point, therefore it is necessary to provide both a rigorous definition and a computation algorithm to formula (\ref{fm}).

The model proposed by Ilinski has a Gaussian Lagrangian, therefore it is defined as
\begin{align}
&P(S(T)|S(0))_{GBM}=\int D\log(S)e^{-\frac{1}{2\sigma^2}\int_{0}^{T}(\partial_t\log(S_t)-r_t)^2dt}\notag\\
&\approx\lim_{N\to\infty}\prod_{i=1}^{N-1}\int_0^\infty \frac{dS_i}{S_i} \exp\bigg[-\frac{T/N}{2\sigma^2}\sum_{i=0}^{N-1}\bigg( \frac{\log(S_{i+1})-\log(S_{i})}{T/N}-r\bigg)^2\bigg]\,.\label{fda}
\end{align}
The Gaussian nature of  (\ref{fda}) guarantees two important properties: the finite dimensional approximation does not depend on $N$ and the probability distribution derived from the calculation is again log-normal.

\begin{figure}[ht]
	\centering
	\includegraphics[width=6.75cm]{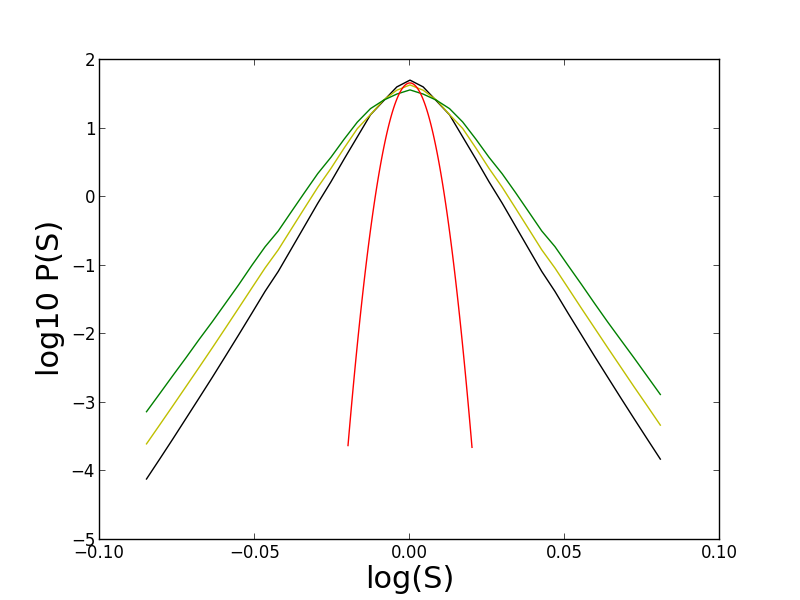}
	\small 
	\caption{Output of the algorithm when $\gamma=0.4$, $p=1.4$ and $\sigma=0.014$, and $N=10$ (blue), $N=11$ (yellow), $N=12$ (green). The red curve represents a Gaussian pdf with $\sigma$=0.00128.}
	\label{Error}
\end{figure}

Consider now (\ref{fm}). The most natural definition would be the analogous of (\ref{fda}), but our computations suggest that the sequence does not approach a finite limit, see Figure \ref{Error}.
In order to propose a well posed definition, we consider first the case $\gamma=1$, that is
\[ P(S(T)|S(0))\propto\int D\log(S)e^{-\frac{1}{2\sigma^p}\int_{0}^{T}|\partial_t\log(S_t)-r_t|^pdt}\,,
\]
and we compare our problem with a standard problem in probability theory.

Consider a collection of i.i.d. stable L\'evy random variables $\{X_n\}$ with  characteristic function
\[\phi(q)=e^{-\gamma|q|^\alpha} \quad \textup{with} \quad\alpha \le 2\,.\]
The previous distribution is stable, see e.g. \cite[Sec 4.2]{M-Sb}; this amounts to say that the sum of $N$
rescaled random variables $\{X_n\}$
\[\tilde{S}_N=\sum_{n=0}^{N}X_n/N^{1/\alpha}\]
is also a  L\'evy random variable with the parameters independent from $N$, provided
that we also rescale the probability $P(\tilde{S})\mapsto P(\tilde{S})N^{1/\alpha}$.

Clearly, when $\alpha=2$ we recover the Gaussian case, \cite[Sec 3.3]{M-Sb}, and the rescaled variables are also Gaussian. 

\begin{figure}[ht]
	\centering
	
    \includegraphics[width=5.75cm]{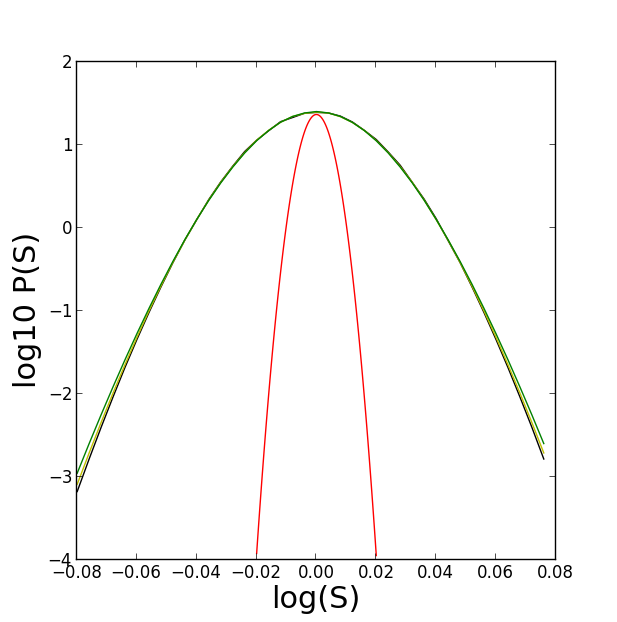}
	\includegraphics[width=5.75cm]{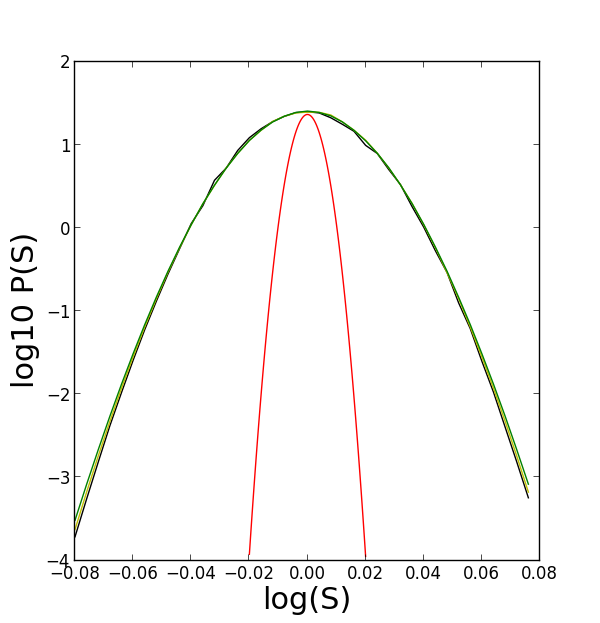}
    \small 
    \caption{Left: results with $\gamma=1$, $p=1.2$, $\sigma=0.0027$, $\Delta \log S=0.16$. Right: results with $\gamma=1$, $p=1.5$ and $\sigma=0.0037$, $\Delta \log S=0.16$. For both pictures we have N=9 (green), N=12 (blue) and N=15 (yellow). The red curve represents a Gaussian distribution with $\sigma$=0.00128.}
    \label{g1p15-2}

\end{figure}

\begin{figure}
		\centering
		\includegraphics[width=5.75cm]{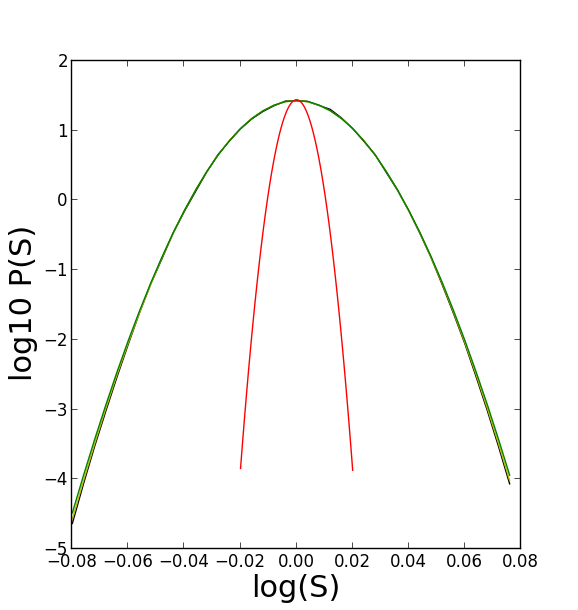}
		\small 
		\caption{Results with $\gamma=1$, $p=1.7$ and $\sigma=0.004$, $\Delta \log S=0.16$ with N=9 (green), N=12 (black) and N=15 (yellow). The red curve represents a Gaussian distribution with $\sigma$=0.00128.}
	    \label{g1p17-e}
	\end{figure}

A similar phenomenon happens with the  path integral definition. We set
\begin{align}
&P(S(T)|S(0))=\int D\log(S)e^{-\frac{1}{2\sigma^2}\int_{0}^{T}|\partial_t\log(S_t)-r_t|^pdt}\notag\\
&\approx\lim_{N\to\infty}\prod_{i=1}^{N-1}\int_0^\infty \frac{dS_i}{S_i} \exp\bigg[-\frac{(T/N)^{p/2}}{2\sigma^p}\sum_{i=0}^{N-1}\bigg| \frac{\log(S_{i+1})-\log(S_{i})}{T/N}-r\bigg|^p\bigg]\,,\label{fda2}
\end{align}
and we check that the numerical approximation appears to be convergent, see Figures \ref{g1p15-2} and \ref{g1p17-e} (left). We recover the familiar quadratic definition in the case $p=2$.  
  
In order to support the choice of the exponent $p/2$, for some choices of $(p,\sigma)$ we estimated the path integrals with different values of $N$, and we observed that the resulting pdf's converge when $N$ is increased.
Figures \ref{g1p15-2} and \ref{g1p17-e} show the results for some of these choices.
The relative errors of the computations with $\gamma=1$ is less than $3\%$ .

	\begin{figure}
		\centering
	    \includegraphics[width=8cm]{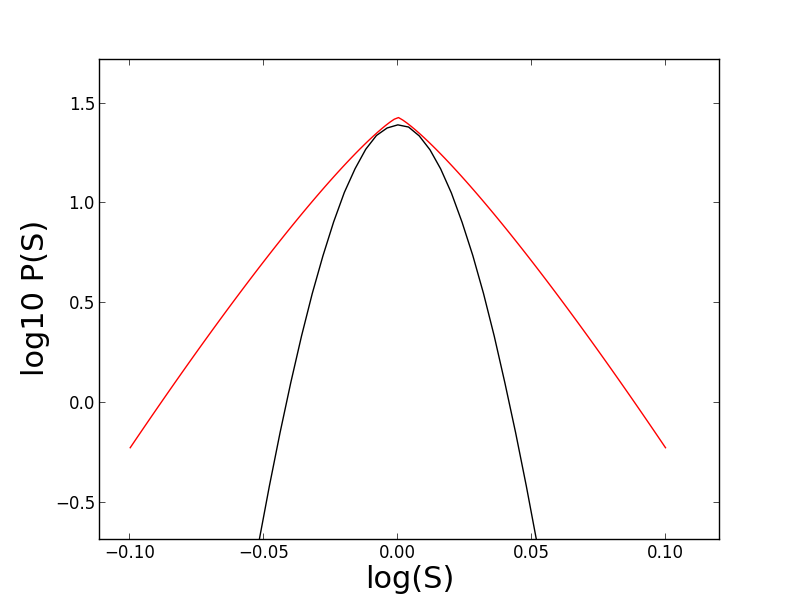}
		\small 
		\caption{Comparison between the graph in Figure \ref{g1p15-2} right (black), and the pdf associated to the action (red) }
	    \label{comp}
	\end{figure}

We point out that the distribution computed with the path integral is different from the one associated to the action. 
\begin{equation*}
\int D\log(S)e^{-\frac{1}{2\sigma^p}\int_{0}^{T}|\partial_t\log(S_t)-r_t|^pdt}
\ne  e^{-\frac{1}{2\sigma^pT}|\log(S_T)-\log(S_0)|^p}\,,
\end{equation*}
see Figure $\ref{comp}$.

\begin{figure}[ht]
		\centering
	    \includegraphics[width=5.75cm]{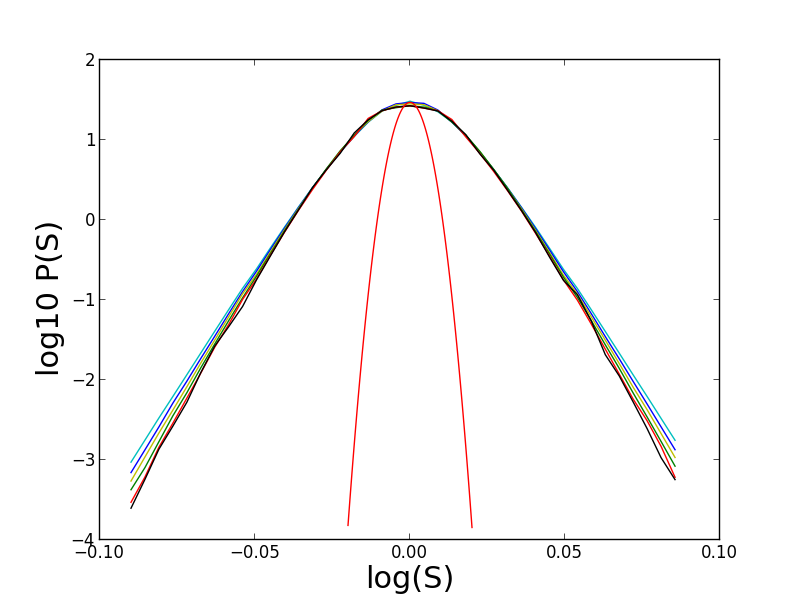}
		\includegraphics[width=5.75cm]{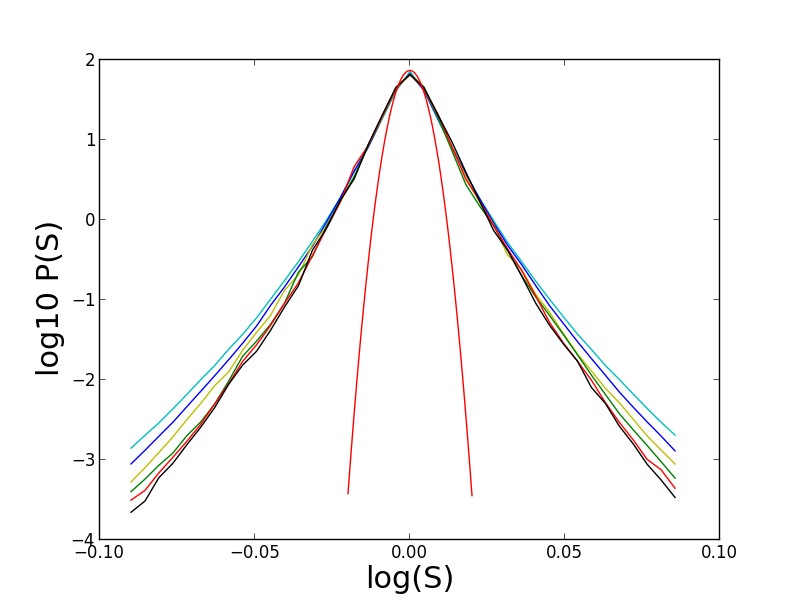}
		\small 
		\caption{Left: results with $\gamma=0.5$, $p=1.5$ and $\sigma=0.015$. Right: results with $\gamma=0.2$, $p=1.3$ and $\sigma=0.014$. The  dimensions are: N=10 (azure), N=11 (blue), N=12 (yellow), N=13 (green), N=14 (red) and  N=15 (black).}
		\label{test5-4}
\end{figure}

\begin{figure}
   \centering
    \includegraphics[width=5.75cm]{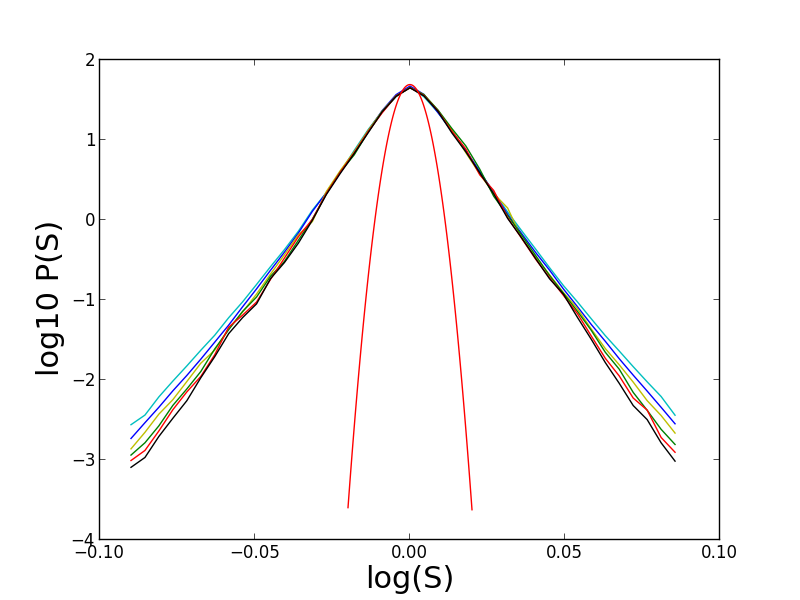}
    \includegraphics[width=5.75cm]{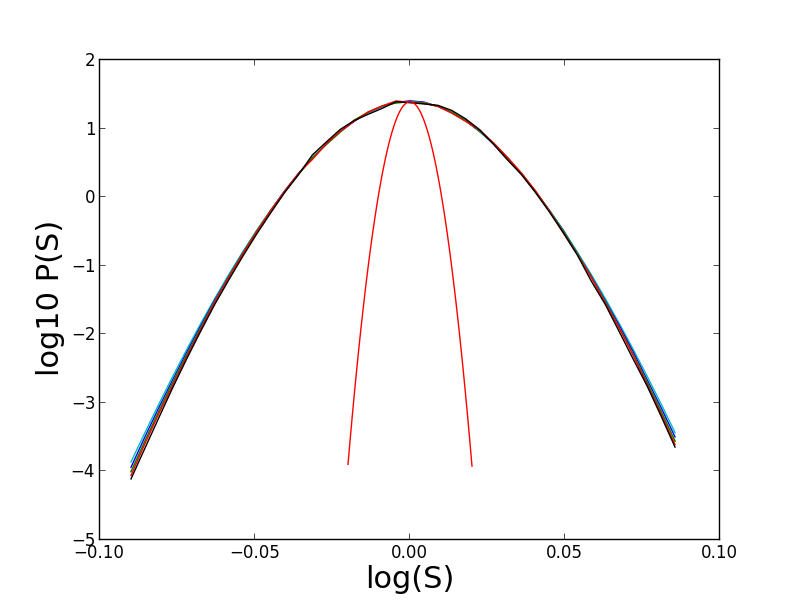}
    \small 
    \caption{Left: results with $\gamma=0.29$, $p=1.275$ and $\sigma=0.013$. Right: results with $\gamma=0.8$, $p=1.7$ and $\sigma=0.0085$. The  dimensions are: N=10 (azure), N=11 (blue), N=12 (yellow), N=13 (green), N=14 (red) and  N=15 (black).}
    \label{test3-2}
\end{figure}

The case with $\gamma \ne 1$ needs some further generalization of the definition of the path integral. With the same approach, we found that a well posed definition is
\begin{align}
&P(S(T)|S(0))=\int D\log(S)e^{-\frac{1}{2\sigma^2}\big(\int_{0}^{T}|\partial_t\log(S_t)-r_t|^pdt\big)^{\gamma}}\notag\\
&\approx\lim_{N\to\infty}\prod_{i=1}^{N-1}\int_0^\infty \frac{dS_i}{S_i} \exp\bigg[-\frac{1}{2\sigma^p}\bigg(\sum_{i=0}^{N-1}\bigg| \frac{\log(S_{i+1})-\log(S_{i})}{T/N}-r\bigg|^p(T/N)^{f(p,\gamma)}\bigg)^{\gamma}\bigg]\,,\label{fda3}
\end{align}
with
\[f(p,\gamma)=p-(p/2)^{\gamma}/\gamma.\]
Figures \ref{test5-4} and \ref{test3-2} show the results of the estimates of the integral for some values of $(\gamma,p,\sigma)$. The relative errors are negligible when $N\le12$. When $N= 13,14,15$ it is respectively $5\%$, $8\%$ and $10\%$.

\begin{figure}[ht]
	\centering
	\includegraphics[width=6.75cm]{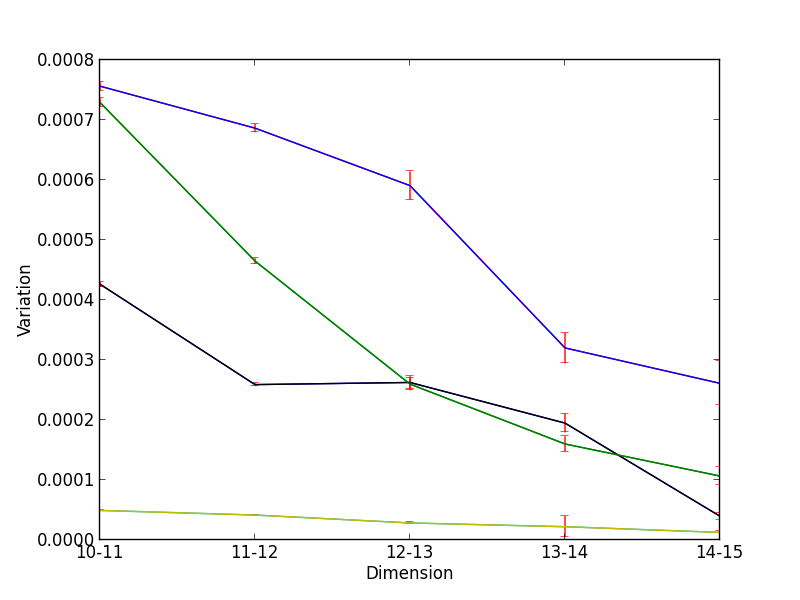}
	\small 
	\caption{The black and green  lines refer to the PDFs in Figure \ref{test5-4}. The blue and yellow lines  refer to the PDFs in Figure  \ref{test3-2}.}
	\label{MP}
\end{figure}

Figures \ref{test3-2} and \ref{test5-4} show the results of our tests concerning the well posedness of (\ref{fda3}).
The probability density functions are studied over a range
of more than 4 orders of magnitude, that is a wider range than what we use in the comparison with real data. In order to estimate the convergence, for a given integer $M$ we compute the approximations with $N=M$ and $N=M+1$; then we compute the average of the differences of the probabilities of the highest and lowest price in the two cases. The results for different choices of $M$ are shown in Figure \ref{MP} along with the relative errors. 
Clearly, we do not have a proof that our definition of non Gaussian path integral is well posed, more precisely that the limits in (\ref{fda2}) and (\ref{fda3}) exist and are finite, but we claim that the results of our computations are stable and consistent with the standard Gaussian case. 

\section{Numerical Methods}\label{sec:nm}
To compute the conditional probability $P(S(T)|S(0))$ we need to estimate multi-dimensional integrals.

The crude Monte Carlo (CMC) method is the simplest to implement, but it is too slow. Markov-Chain Monte Carlo (MCMC) methods are faster, but they require a separate computation to normalize the pdf, thus making these methods less efficient for our purpose.

We choose instead the Quasi Monte Carlo (QMC) method.
The main difference between the QMC and CMC consists in the disposition of the generated points.
The QMC algorithm can generate a pattern of points which are distributed more evenly than the plain CMC.
The better distribution of the points in the QMC yields a convergence rate of the integral of order $\log(N)^{D}/N$, where $N$ is the number of points and $D$ the dimension of the integral, while the CMC has order $N^{-1/2}$. Clearly, the QMC method a better choice when $D$ is small, while CMC performs better in higher dimensions. Due to this differences all the simulations shown in section 5 are performed with a QMC methods; whereas the simulation shown in section 6, which are characterized by higher dimensions, are carried on with the usual CMC methods.

There are many algorithms to generate a set of QMC points. We choose  Sobol, because it ensures the same homogeneity of the points in all the directions of the domain.

The QMC algorithm is deterministic method are characterized by a fixed disposition; as opposed  with the quasi-random points generated by the CMC algorithm, therefore errors cannot be estimated by repeating the algorithm. Instead, to obtain different estimates of the integral we shift the points by a random value and then repeat the computation. Details about the Sobol algorithm and the random shift method to estimate the error can be found in $\cite{Gla}$.

We estimated all the probability density functions with 40 final prices. On average the computation required a few minutes in the cases treated with the QMC, using on a laptop with a current dual core processor. We remark that the algorithm is fully parallelizable; in particular the computation times on a GPU may be much smaller.

\section{The Chapman-Kolmogorov equation}\label{sec:ck}
\begin{figure}[ht]
	\centering
	\includegraphics[width=5.75cm]{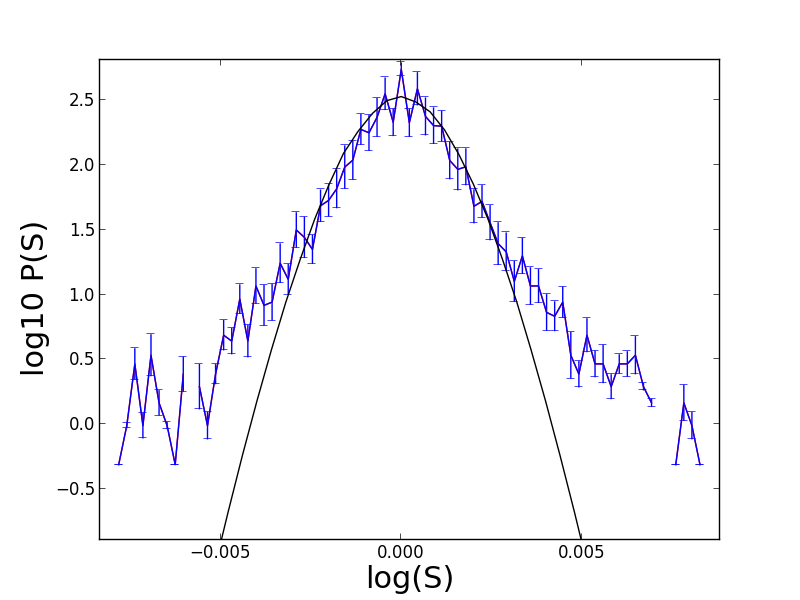}
	\small 
	\caption{  Output of the algorithm  with $\gamma=1$, $p=1.2$ and $\sigma=0.00037$.}
	\label{ndp}
\end{figure}

The Geometric Brownian Motion is a Markov process. This means that the probability density function obtained from the path integral formulation satisfies the gauge-invariant semigroup equation
\begin{equation}\label{ck}
P(S(T)|S(0))_{GBM}
=\int D\log(S_K)P(S(T)|S(K))_{GBM}P(S(K)|S(0))_{GBM}\,,
\end{equation}
where $K \in (0,T)$.\\
The previous equation is also known as the Chapman-Kolmogorov equation and it is a general property of  Markovian processes. A straightforward calculation shows that (\ref{ck}) is satisfied if and only if $\gamma=1$.  In  \cite[Sec. 20.1.18]{HKl}, it is stated that the semigroup property is reasonably satisfied for  time frames higher than 15 minutes only outside the deep tails of rare events. Also, the model proposed for the one minute dynamics, that is the Boltzmann distribution, satisfies the  semigroup equation, but it does not achieve a good agreement in the deep tails region, see \cite[Figure 20.10]{HKl}. The model in (\ref{fm}) behaves in the same manner; Figure \ref{ndp} compares the results of the model with $\gamma=1$ with the $GE$ data shown in Figure \ref{g5}. The agreement is satisfactory only in the central region.

The numerical computation shown in Section 6  are based on a model which  does not satisfies  equation (\ref{ck}); indeed this model shows a good agreement also in the deep tails region.

We observe that the computation with $\gamma\simeq 1$, see Figures \ref{g1p15-2} (left),\ref{g1p17-e}, \ref{test5-4} (left),\ref{test3-2} (right)  have good overlap through all the finite dimensional approximations.
The computations with $\gamma <0.4$, see
Figures   \ref{test5-4} (right) and \ref{test3-2} (left), have a good overlap only in the higher dimensional approximations;
and the differences between the extreme prices are higher with respect the cases with $\gamma$ closer to 1.
This suggests that the parameter $\gamma$ plays an active role in the speed of convergence for the finite dimensional approximations.

\section{Conclusion}
In this paper we generalize the model proposed by Ilinski by using a Lagrangian which is not a function of the arbitrage. This generalization allows us to define a new model, where the main dynamics is non-Gaussian. The main advantage of this model consists in the low number of parameters with respect to other models with similar precision. More precisely, our model is characterized by 3 parameters, whereas the models described in \cite{D-F},\cite{PA} and the Heston model have 5 parameters.

The application of a non-Gaussian dynamics to financial problems has already been proposed in \cite{HKla1,HKla2,HKla3,HKl}. Our model is also characterized by the  fact that it does not satisfy the semigroup property, i.e. the stochastic process described by the path integral, is not Markovian.
Path integrals have already been used in order to solve stochastic differential equation with Non-Markovian noise, see e.g. \cite{NMPI1,NMPI2,NMPI3}.
Our work adds to the vast literature in this field and proposes a new conjecture for the computation of a particular form of path integrals.
Our result are in agreement with the financial data analysed and with the theoretical background proposed.

Some interesting questions remain open.
We analyse the prices of 3 stocks and 2 indexes with different time horizons, but we do not study the auto-aggregation phenomena because our dataset is too small for this kind of analysis.
The study of a  connection between the auto-aggregation constant $\alpha$ and the parameters $(\gamma,p)$ is in our opinion worth investigating. 
Another interesting point may be the fact that the optimal value of $\gamma$ seems to be $p-1$.\\
Finally, we think that it would be important to study the connection between the parameter $\gamma\ne1$ associated to the presence of memory and the relative implications with the convergence of the finite dimensional approximations.
\bigskip

{\bf Acknowledgement.} The first author is grateful to the ETH Z\"urich for the kind hospitality, and in particular to Josef Teichmann for fruitful discussions.

\end{document}